\documentstyle[preprint,pre,aps]{revtex}
\draft

\input{psfig}
\begin{document}


\title{Square Patterns and Quasi-patterns in Weakly Damped Faraday Waves}
\author{Wenbin Zhang$^1$ and Jorge Vi\~nals$^2$}
\address{$^1$Department of Chemical Engineering, Massachusetts Institute of
Technology, Cambridge, Massachusetts 02139 \\
$^2$Supercomputer Computations Research Institute,
Florida State University, Tallahassee, Florida 32306-4052, \\
and Department of Chemical Engineering,
FAMU/FSU College of Engineering, Tallahassee, Florida 31316}

\date{\today}
\maketitle

\begin{abstract}

Pattern formation in parametric surface waves is studied in the limit of
weak viscous dissipation. A set of quasi-potential equations (QPEs) is 
introduced that admits a closed representation in terms of surface variables 
alone.  A multiscale expansion of the QPEs reveals the importance of triad
resonant interactions, and the saturating effect of the driving force
leading to a gradient amplitude equation. Minimization of the associated
Lyapunov function yields standing wave patterns of square symmetry for
capillary waves, and hexagonal patterns and a sequence of quasi-patterns 
for mixed capillary-gravity waves. Numerical integration 
of the QPEs reveals a quasi-pattern of eight-fold symmetry in the range of
parameters predicted by the multiscale expansion.

\end{abstract}



When a fluid layer with a free upper surface is subjected to
vertical oscillation, Faraday waves are observed \cite{re:miles90,re:cross93}. 
In a large enough container and for low viscosity fluids, standing 
wave patterns of square symmetry are observed near threshold 
\cite{re:lang62}. Based on amplitude equations that we derive below, 
the experimental observation of square patterns in the capillary-dominated
regime is explained.  We also predict that hexagonal 
and a sequence of quasi-patterns can be stabilized 
for the case of a sinusoidal driving force as a result of triad resonant 
interactions for mixed capillary-gravity waves \cite{re:edwards95}.
To our knowledge, this is the first theoretical derivation that starting
from a realistic model of the fluid shows that a quasi-crystalline pattern
is a stable steady state, and corroborates the conjecture of Newell and
Pomeau \cite{re:newell93b} on the existence of the so-called \lq\lq turbulent 
crystals".

Pattern-forming instabilities occur in a variety of extended nonlinear
systems. The emergence of spatial patterns close to onset of the 
instability can often be described by amplitude equations 
\cite{re:cross93,re:newell69}.
However, for near-Hamiltonian (or weakly dissipative) systems, there is no
general agreement on how dissipation 
should be incorporated into the amplitude equation formulation.
Previous work on Faraday waves was based on amplitude equations for
a purely Hamiltonian system, 
to which linear and nonlinear damping terms were added
by introducing a dissipation function \cite{re:milner91,re:miles93}. 
In this approach, linear dissipative effects in the original system
contribute only to linear damping terms in the associated amplitude equations, 
while nonlinear damping terms result entirely from nonlinear dissipative 
effects. Such an approach has contributed to the general belief that for 
near-Hamiltonian systems, nonlinear saturation of the linear instability 
does not occur if only linear dissipative effects are considered, and weak 
nonlinear dissipative or other higher order effects are needed 
for nonlinear saturation \cite{re:cross93,re:coullet94}. 
In this paper, we show that in the case of weakly damped 
parametric surface waves linear dissipative 
effects do contribute to the nonlinear damping terms in the amplitude 
equation, and that they {\em alone} can saturate the parametric instability.
In addition, the experimental observation of square patterns in 
capillary-dominated regime is naturally explained without having to invoke 
poorly understood nonlinear dissipative effects, or higher order terms in 
the amplitude equation.

The basic difference with previous studies \cite{re:milner91,re:miles93} 
is that although the bulk flow does remain 
potential, it is modified by a rotational viscous boundary layer near 
the free surface that has to be explicitly incorporated into the
analysis \cite{re:lundgren88,re:ruvinsky91,re:zhang94}.
When the thickness of the viscous boundary layer is small compared to
the typical wavelength of the pattern, the weak effects due to viscosity
can be taken into account by introducing effective boundary conditions
for the otherwise potential bulk flow.  This is the basic idea of 
the quasi-potential approximation introduced below.
We first expand the equations governing the motion of an incompressible 
viscous fluid and the appropriate boundary conditions at the free surface 
in the small thickness of the free surface boundary layer, $\delta$. 
The resulting equations 
are further simplified by recasting them in a nonlocal form that involves 
the flow variables on the free surface only; thus eliminating the 
need to explicitly solve for the flow in the bulk \cite{re:zhang94}.
Let $z$ be the normal direction to the surface at rest and
$g(t) = -g_0 - g_{z0} \sin \Omega t$ the driving force where $g_0$ is 
the constant acceleration of gravity, and $\Omega$ and $g_{z0}$ are the 
angular frequency and the amplitude of the driving force respectively. 
We choose $1/\omega_0 \equiv 2/\Omega$ as the unit of time 
and $1/k_0$ as the unit of length, with $k_0$ defined by
$ \omega_0^2 = g_0 k_0 + \Gamma k_0^3/\rho$, the linear
dispersion relation for surface waves, where $\Gamma$ is the surface 
tension and $\rho$ the density of the fluid.  Then the dimensionless, 
nonlocal and quasi-potential equations read \cite{re:zhang94},
\begin{eqnarray}
\label{eq:QPEh}
\partial_t h({\bf x},t)  = \gamma\nabla^2 h + {\hat{{\cal D}}}\Phi 
    - \nabla \cdot (h\nabla\Phi)
    + \frac{1}{2} \nabla^2 (h^2{\hat{{\cal D}}}\Phi)
\nonumber \\
 -{\hat{{\cal D}}}(h{\hat{{\cal D}}}\Phi) 
+ {\hat{{\cal D}}}\left[h{\hat{{\cal D}}}(h{\hat{{\cal D}}}\Phi) 
+ \frac{1}{2}h^2\nabla^2\Phi\right], \\
\label{eq:QPEPhi}
\partial_t \Phi({\bf x},t) = \gamma\nabla^2\Phi 
                       - \left(G_0-\Gamma_0\nabla^2\right) h - 4fh \sin 2t
\nonumber \\
     +\frac{1}{2}({\hat{{\cal D}}}\Phi)^2 
     -({\hat{{\cal D}}}\Phi)\left[h\nabla^2\Phi 
     + {\hat{{\cal D}}}(h{\hat{{\cal D}}}\Phi)\right]
\nonumber \\
     -\frac{1}{2}(\nabla\Phi)^2
     -\frac{\Gamma_0}{2}\nabla\cdot\left(\nabla h(\nabla h)^2\right),
\end{eqnarray}
where ${\bf x}\!=\!(x,y)$, $\nabla\!=\!(\partial_x,\partial_y)$, $h$ is the 
surface displacement away from planarity, $\Phi$ is the 
value of velocity potential at the free surface (the surface velocity
potential), $\gamma = 2\nu k_0^2/\omega_0$ is a 
linear damping coefficient ($\gamma \sim \delta^2 \ll 1$), 
$G_0 = g_0k_0/\omega_0^2$, $\Gamma_0 = \Gamma k_0^3/(\rho\omega_0^2)$, 
and $f = g_{z0}k_0/(4\omega^2_0)$ is the driving
amplitude, and ${\hat{{\cal D}}}$ is a Fourier-integral operator, which 
is defined for an 
arbitrary function $u({\bf x})$ by ${\hat{{\cal D}}} u({\bf x}) = 
\int_{-\infty}^{\infty}|{\bf k}|\hat{u}({\bf k}) \exp(i{\bf k}\cdot{\bf x})
d{\bf k}$ with $\hat{u}({\bf k})$ the Fourier transform of $u({\bf x})$.
Notice that $G_0 + \Gamma_0 = 1$ by definition.
Only those viscous terms that are linear in the surface variables are
retained in Eqs.~(\ref{eq:QPEh}) and (\ref{eq:QPEPhi}).  Nonlinear viscous
terms, which were referred to as nonlinear dissipative effects in the 
introduction, have been neglected. 
Both the asymptotic analysis that follows and the
numerical results presented later are based on Eqs.~(\ref{eq:QPEh}) and
(\ref{eq:QPEPhi}).

Linear analysis indicates that for $\gamma\ll 1$ the 
planar surface becomes unstable at 
$f_{c}=\gamma$, and the critical wavenumber is $k_0=1$.  
We seek nonlinear standing wave solutions near threshold 
($\varepsilon \equiv (f-\gamma)/\gamma \ll 1$) and expand 
Eqs.~(\ref{eq:QPEh}) and (\ref{eq:QPEPhi}) in $\varepsilon^{1/2}$ with 
multiple time scales,
$h({\bf x}, t, T) = \varepsilon^{1/2} h_1({\bf x}, t, T) + \varepsilon h_2 
+ \varepsilon^{3/2} h_3 + \cdots $ and
$\Phi({\bf x}, t, T) = \varepsilon^{1/2} \Phi_1({\bf x}, t, T) + \varepsilon 
\Phi_2 + \varepsilon^{3/2} \Phi_3 + \cdots$, where $T=\varepsilon t$.
At ${\cal O}(\varepsilon^{1/2})$, we consider a set 
of $N$ standing wave modes with critical wavevectors $\pm{\hat{k}_j}$; then 
the linear solution reads,
\begin{eqnarray}
\label{eq:order1_h}
h_1 &\!=\!& \left(\cos t + \frac{f}{4} \sin 3t\right)
              \!\sum_{j=1}^{N}\biggl[A_j(T) \exp(i{\hat{k}_j}\!\cdot\!{\bf x})
                                + c.c.\biggr], \!\! \\
\label{eq:order1_Phi}
\Phi_1 &\!=\!&
     \left(-\sin t + f\cos t + \frac{3f}{4}\cos 3t\right)
\nonumber \\
  && \hspace*{2cm} \times 
     \sum_{j=1}^{N}\biggl[A_j(T)\exp(i{\hat{k}_j}\!\cdot\!{\bf x})+c.c.\biggr],
\end{eqnarray}
where the complex amplitudes $A_j$ are assumed to vary in the slow time
scale $T$.  Notice that the linear solution contains not only the 
subharmonic responses but also the terms that are proportional to the 
driving force $f$.  These latter terms 
arise because $f$ is finite for the expansion in $\varepsilon$.  
Terms that are of higher order in $f$ have been neglected.  
In what follows, we shall replace $f$ in the linear solution by $\gamma$ 
since $f = \gamma (1+\varepsilon)$.

At ${\cal O}(\varepsilon)$, there is no solvability condition; however 
there are resonant interactions that have to be taken into account.
A particular solution for $h_2$ can be written as
\begin{eqnarray}
h_2 = \sum_{j,l=1}^{N}\biggl\{
    H_{jl}(t)\left[A_jA_l\exp\left(i({\hat{k}_j}+{\hat{k}_l})\cdot{\bf 
x}\right) + c.c.\right]
\nonumber \\
   +H_{j,-l}(t)\left[A_jA_l^*\exp\left(i({\hat{k}_j}-{\hat{k}_l})
                                       \cdot{\bf x}\right)+ c.c.\right]
\biggl\},
\end{eqnarray}
where the $H_{jl}(t)$ are unknown functions that satisfy,
\begin{eqnarray}
\partial_{tt}H_{jl} + 2\gamma\sqrt{2(1+c_{jl})}\partial_t H_{jl}
+ [G_0+2\Gamma_0(1+c_{jl})]
\nonumber \\
\times \sqrt{2(1+c_{jl})}H_{jl} 
\label{eq:resonance_explained}
= F_{jl}^{(1)}\cos 2t + F_{jl}^{(2)}\sin 2t + \cdots,
\end{eqnarray}
where $c_{ij} \equiv \cos \theta_{ij} = \hat{k}_{j} \cdot \hat{k}_{l}$, and
$F_{jl}^{(1)}$ and $F_{jl}^{(2)}$ are proportional to  the amplitudes
$A_jA_l$.  Equation (\ref{eq:resonance_explained}) 
is the equation of an additively forced harmonic oscillator 
with friction. When the ``natural" frequency of the ``oscillator" equals 
to the driving frequency, resonance occurs.  This condition reads,
$[G_0+2\Gamma_0(1+c_{jl})]\sqrt{2(1+c_{jl})} = 4$.  Due to the nonzero 
damping coefficient, this resonance results in a finite value for 
$H_{jl}$ that is inversely proportional to the damping coefficient.
Since the RHS of Eq.~(\ref{eq:resonance_explained}) is proportional to 
$A_jA_l$, there are three waves involved in this resonance, namely,
standing wave modes $A_j$ and $A_l$, and mode $B$ with wavevector
${\hat{k}_j}+{\hat{k}_l}$.  Therefore, 
Eq.~(\ref{eq:resonance_explained}) describes a three-wave resonant interaction.
Note that the wavenumber for mode $B$ is away from the critical wavenumber
$k_0=1$; thus mode $B$ is a linearly stable mode.  
The relevance of triad resonant interactions to pattern selection can be
understood intuitively.  Since the resonant growth of linearly damped mode
$B$ is at the expense of the reduction of the amplitudes $A_j$ and $A_l$,
the growth of modes $A_j$ and $A_l$ are less favored than any other
mode $A_m$ that does not participate in a triad resonant interaction.
In other words, the system tries to avoid critical modes that participate in  
triad resonant interactions.  As shown below, triad resonant interactions 
strongly influence pattern selection through coefficients of cubic nonlinear
terms in the amplitude equations in a way that is consistent with the above 
argument.  Finally, for $\Gamma_0 < 1/3$, triad resonance is not possible.  
For $\Gamma_0=1/3$, wavevectors of the three resonating waves are in the 
same direction ($\theta_{jl}^{(r)} = 0$).  As $\Gamma_0$ is further increased, 
$\theta_{jl}^{(r)}$ also increases.  For purely capillary waves 
($\Gamma_0 = 1$), $\theta_{jl}^{(r)}$ reaches the maximum value of 
{$\theta_{jl}^{(r)} \approx 74.9^{\circ}$ or $c_{jl} = 2^{1/3}-1$.

At ${\cal O}(\varepsilon^{3/2})$, we obtain the standing wave amplitude
equations (SWAEs) from a non-trivial solvability condition,
\begin{equation}
\label{eq:swe}
\frac{\partial A_j}{\partial T} =  \gamma A_j - \biggl[\gamma g(1)|A_j|^2
     +\gamma\!\!\!\sum_{l=1(l\ne j)}^{N}\!\! g(c_{jl})|A_l|^2\biggr] A_j,
\end{equation}
where $j=1,2, \cdots, N$, and $g(1)$ and $g(c_{jl})$ are given in 
Ref.~\cite{re:zhang94}.  There are two kinds of contributions to 
$g(1)$ and $g(c_{jl})$.  One is from the linear viscous terms 
in Eqs.~(\ref{eq:QPEh}) and (\ref{eq:QPEPhi}).  The other is due to the 
parametric driving force and proportional to the driving amplitude $f$.  
These two kinds of contributions appear together in Eq.~(\ref{eq:swe}) 
since we have set $f=\gamma$ at the linear order.
The latter contribution is directly related to the terms  
proportional to $f$ in the linear solution 
(Eqs.~(\ref{eq:order1_h}) and (\ref{eq:order1_Phi})), and it provides an 
{\em amplitude-limiting effect}. 
The nonlinear interactions between the primary subharmonic modes and
terms related to the driving force in the linear solution produce terms 
that are out of phase by $\pi/2$ with the primary subharmonic mode, 
and thus can contribute to saturate the wave amplitude.  An important
point is that this amplitude-limiting effect results from the forcing
term, but not from a dissipative term. As a result, this effect is also
important even for Hamiltonian systems.

It can be shown that $g(1)>0$ \cite{re:zhang94}, which indicates the 
bifurcation to the standing wave state is supercritical.  We rescale 
the amplitudes as $\sqrt{g(1)}A_j$ and the coefficients as $g(c_{jl})/g(1)$.
In what follows, we shall only refer to the scaled amplitudes and the
scaled coefficients; but use the same notation for them as for the 
unscaled ones. 
Note that the scaled coefficient $g(c_{jl}\rightarrow\pm 1)=2$.
Figure \ref{fi:gtheta}(a) and (b)
shows the scaled function $g(c_{jl})$ for two different values
of the damping coefficient $\gamma$ and $\Gamma_0=1$.  The maxima 
in $g(c_{jl})$ around $c_{jl}=0.26$ ($\theta_{jl} = 74.9^{\circ}$) 
correspond to the triad resonance for purely capillary waves.
The function $g(c_{jl})$ for capillary-gravity 
waves of $\Gamma_0 = 1/3$ is shown in Fig.~\ref{fi:gtheta}(c) and (d).
Since the triad resonant interaction occurs among waves with their wavevectors
in the same direction when $\Gamma_0 = 1/3$, the resonant peaks (or maxima)
are at $c_{jl}=\pm 1$.  Instead of large peaks at $c_{jl}=\pm 1$, we see 
that $g(c_{jl})$ has a wide flat region around $c_{jl}=0$ and reaches very 
small positive values for small $\gamma$ due to scaling.
In all cases, the effect of triad resonance is weaker for larger
values of $\gamma$ as expected.

\begin{figure}[hbt]
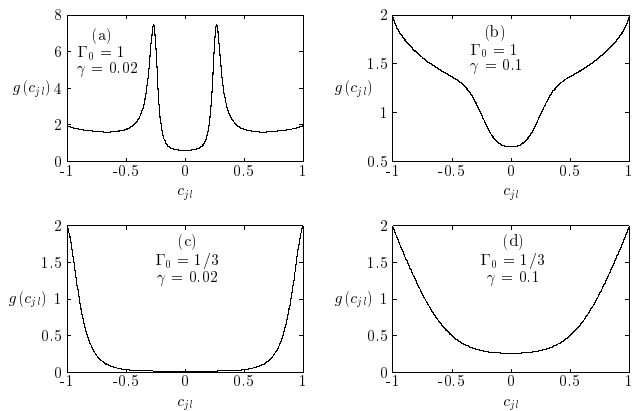

\hbox{\psfig{figure=wz3.fig1a,width=1.65in}
      \psfig{figure=wz3.fig1b,width=1.65in}}
\hbox{\psfig{figure=wz3.fig1c,width=1.65in}
      \psfig{figure=wz3.fig1d,width=1.65in}}
\caption{The coefficient $g(c_{jl})$ as a function of $c_{jl}$
for purely capillary waves ($\Gamma_0=1$) with the linear damping
coefficient $\gamma=0.02$ (a), and $\gamma=0.1$ (b). The same
coefficient for gravity-capillary waves of $\Gamma_0=1/3$ with
$\gamma=0.02$ (c) and $\gamma=0.1$ (d).}
\label{fi:gtheta}
\end{figure}

The issue of pattern selection can be discussed by noting that
Eq.~(\ref{eq:swe}) is of gradient form 
$1/\gamma \partial_T A_j = -\partial {\cal F}/\partial A_j^*$.
Apart from the trivial solution of $A_j =0$ for $j=1, \cdots, N$,
Eq.~(\ref{eq:swe}) has a family of stationary solutions differing in the 
total number of standing waves $N$ for which $A_j \ne 0$.  By considering
the case in which the magnitudes of all standing waves are the same, 
Eq.~(\ref{eq:swe}) has the following solutions,
$|A_j| = |A| = \left(1+\sum_{l=1(l\ne j)}^{N} g(c_{jl})\right)^{-1/2}$.
The Lyapunov function for these solutions are,
${\cal F} = -\frac{N}{2}|A|^2 = -\frac{N}{2}/[1+\sum_{l=1(l\ne j)}^{N} 
g(c_{jl})]$.
For $N=1$ (parallel roll solution), ${\cal F}_1 = -\frac{1}{2}$.  
For $N=2$, we have either square ($c_{12} = 0$) or rhombic ($c_{12} \ne 0$) 
patterns with ${\cal F}_2 = -1/(1+g(c_{12}))$.
If we consider only regular patterns, {\em i.e.}~pattern structures for 
which the angle between any two adjacent wavevectors ${\bf k}_j$ and ${\bf 
k}_{j+1}$ is the same and amounts to $\pi/N$, we have either hexagonal or 
triangular patterns for $N=3$. 
Regular patterns for $N \ge 4$ are two-dimensional quasi-crystalline patterns 
(or quasi-patterns \cite{re:edwards94}).  A quasi-pattern has long-range
orientational order but no spatial periodicity, thus analogous to a 
quasi-crystal in solid state physics.  
Such patterns have been already observed in experiments of Faraday waves
in systems driven by {\em two} carefully chosen frequency components
\cite{re:edwards94}, but not in the single frequency case analyzed here.
Fig.~\ref{fi:lyapunov} shows the Lyapunov function ${\cal F}_N$ as a function 
of $\gamma$ for $N=1,2,3,4,5,6,7,8$, and $\Gamma_0 = 1/3$ and 1.  
For $\Gamma_0 = 1$, patterns of square symmetry ($N=2$) have the
lowest values of ${\cal F}_N$ for all values of $\gamma < 0.2$, in agreement
with experiments. However, patterns of different symmetries are favored in 
different ranges of $\gamma$ for $\Gamma_0 = 1/3$. 
This can be understood qualitatively by noting that the self-interactions
of the critical modes are less favored than interactions among them.  
Consequently, pattern structures with large $N$ are favored.

\begin{figure}[hbt]
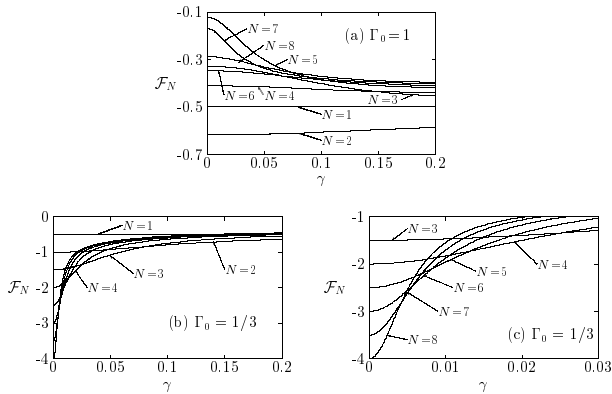

\hbox{\hspace*{0.75in}
      \psfig{figure=wz3.fig2a,width=1.6in}}
\hbox{\psfig{figure=wz3.fig2b,width=1.6in}
      \psfig{figure=wz3.fig2c,width=1.6in}}
\caption{The values of the Lyapunov function ${\cal F}_N$ ($N=1,2,3,4,5,6,7,8$)
as a function of $\gamma$ for purely capillary waves ($\Gamma_0 = 1$)
in (a), and for capillary-gravity waves of $\Gamma_0 =1/3$ in (b).
Part of (b) with small $\gamma$ is shown in (c).}
\label{fi:lyapunov}
\end{figure}

We next present results from numerical solutions of the
QPEs (Eqs.~(\ref{eq:QPEh}) and (\ref{eq:QPEPhi}))
to check the stability of the
quasi-patterns described above. An analytic stability calculation is far
too involved. We use a Fourier-Galerkin 
spectral method \cite{re:canuto86} with periodic boundary conditions in a 
square domain. Time
discretization is of second order.  We use the trapezoidal scheme for the 
linear terms and the second order Adams-Bashforth scheme for nonlinear terms. 
The nonlinear terms are calculated by a pseudospectral method by using Fast 
Fourier Transforms.  Most of the numerical studies are done for a system size
of $64\pi \times 64\pi$, or $32\lambda_0 \times 32\lambda_0$ with the critical 
wavelength $\lambda_0 = 2\pi/k_0 = 2\pi$, and a total number of 256 Fourier 
modes is used for each axis.  We use a time step $\Delta t = 0.04 - 0.1$.
Extensive numerical studies have been performed for different values of the 
three dimensionless parameters $\Gamma_0$, $\gamma$, and $\varepsilon$ in the 
QPEs.  

\begin{figure}[hbt]
\caption{Configuration of $h({\bf x},t)$ at $t=32000$ is shown in gray scale
for $\gamma=0.02$, $\Gamma_0 = 1/3$, and $\varepsilon = 0.1$.}
\label{fi:qpattern}
\end{figure}

In the case of purely capillary waves ($\Gamma_0=1$),
Eqs.~(\ref{eq:QPEh}) and (\ref{eq:QPEPhi}) are integrated 
for $\gamma=0.02, 0.05, 0.1$, 0.2, and small values 
of $\varepsilon = 0.02, 0.05$, 0.1, from a random initial 
condition (gaussianly distributed with zero mean and a variance of $10^{-4}$) 
for field $h({\bf x},t=0)$, and zero values for $\Phi({\bf x},t=0)$. 
Asymptotically regular patterns of standing waves with square symmetry are
obtained for all the above parameters; thus verifying the results of the
asymptotic analysis presented earlier, and again in agreement with
experimental observation.  For capillary-gravity waves at
$\Gamma_0 = 1/3$, the QPEs are integrated for three different values of 
$\gamma = 0.02, 0.05$, and 0.1 at $\varepsilon=0.05$, with the same initial
conditions.  For $\gamma=0.1$, the 
long time configuration of $h({\bf x},t)$ is of approximate square symmetry.   
For $\gamma=0.05$, the long time configuration of $h({\bf x},t)$ is of 
approximate hexagonal symmetry. For $\gamma=0.02$, and to avoid
finite size effects (expected to be stronger for smaller damping 
coefficient $\gamma$ \cite{re:zhang94,re:edwards94}),
we have performed numerical studies for a system of 
size $64\lambda_0 \times 64\lambda_0$ with 512 Fourier modes for each 
axis and a time step $\Delta t=0.1$. The initial condition for $h$ is a 
set of gaussianly distributed random numbers with zero mean and a variance
of $10^{-6}$, and $\Phi$ is set to zero initially.  The configuration 
of $h({\bf x},t)$ at $t=32000$ is shown in Fig.~\ref{fi:qpattern}.
The structure factor of this configuration has been computed
and has eight peaks, which correspond to eightfold 
symmetry of the standing wave pattern. This can also be seen by viewing 
Fig.~\ref{fi:qpattern} at a glancing angle. Thus, the basic prediction 
of the SWAEs is confirmed.  
Finally, we remark that the possible confirmation of stable pattern 
structures of even lower symmetry that are predicted by the SWAEs for
smaller values of the damping coefficient $\gamma$ would require systems 
of larger size.

This work was supported by the Microgravity Science and Application 
Division of the NASA under Contract No. NAG3-1284.
This work was also supported in part by the Supercomputer
Computations Research Institute, which is partially funded by the U.S.
Department of Energy, contract No. DE-FC05-85ER25000.

\bibliographystyle{prsty}
\bibliography{references}

\end{document}